\documentclass[twocolumn]{aastex701}
\usepackage{bm}

\begin{document}

\title{Formation and disruption of resonant chains of super-Earths: Secular perturbations from outer eccentric embryos}

\author[0000-0002-8300-7990]{Masahiro Ogihara}\email{ogihara@sjtu.edu.cn}
\affiliation{Tsung-Dao Lee Institute, Shanghai Jiao Tong University, 1 Lisuo Road, Shanghai 201210, China}
\affiliation{School of Physics and Astronomy, Shanghai Jiao Tong University, 800 Dongchuan Road, Shanghai 200240, China}

\author[0000-0002-1932-3358]{Masanobu Kunitomo}\email{kunitomo.masanobu@gmail.com}
\affiliation{Department of Physics, Kurume University, 67 Asahimachi, Kurume, Fukuoka 830-0011, Japan}

\begin{abstract}

Recent observations have revealed the distribution of orbital period ratios of adjacent planets in multiple super-Earth systems and how these distributions change with time.  
The aim of this study is to clarify under what conditions the observed features of orbital period ratios of super-Earths can be explained, and to identify what causes the dynamical instability of super-Earths captured into resonant chains.  
We perform $N$-body simulations for 100 Myr that follow the formation and orbital evolution of super-Earths originating from a ring of planetary embryos at 1 au from the star.  
The simulations show that super-Earths undergo inward migration in the disk and are captured into mean-motion resonances with their neighbors. As a result, several resonant pairs form a resonant chain. After disk dispersal, some of these chains become dynamically unstable. In such cases, the final distribution of orbital period ratios and their time evolution can be consistent with recent observations. The instabilities of resonant chains are likely triggered by secular perturbations from embryos that remain on outer orbits beyond 1 au, indicating that not only giant planets but also small embryos can disrupt the resonances among inner super-Earths.
We therefore further investigate the secular perturbations from outer embryos using analytic formulas and additional orbital calculations.  
We discuss the conditions required to excite the eccentricities of inner super-Earths on a timescale of about 100 Myr. These conditions include the need for large eccentricities of the outer embryos, as well as constraints on their masses and semimajor axes.

\end{abstract}



\section{Introduction}\label{sec:intro}

Since the discoveries by the {\it Kepler} mission, many properties of close-in super-Earths have been revealed \citep[e.g.,][]{2023ASPC..534..839L, 2023ASPC..534..863W}. In particular, we now have information on their size distribution, orbital distribution, orbital spacing, eccentricities, and system multiplicities. These observed characteristics provide valuable constraints on the origin and formation of super-Earths.

Previous simulations of super-Earth formation have emphasized the importance of reproducing the distribution of orbital period ratios among neighboring planets in multi-planet systems \citep[e.g.,][]{2013ApJ...775...53H, 2015A&A...578A..36O, 2017MNRAS.470.1750I, 2019A&A...627A..83L}. A key feature found by {\it Kepler} is that most adjacent super-Earth pairs are not located in mean-motion resonances (MMRs). This feature has been investigated as a possible outcome of long-term dynamical evolution, in which super-Earths initially captured into resonant chains during the gas disk phase later lose their commensurabilities \citep[e.g.,][]{2018A&A...615A..63O, 2021A&A...650A.152I, 2024A&A...692A.246B, 2025ApJ...979L..23S}.

While the {\it Kepler} sample consists mainly of relatively old stars (median age $\sim$4 Gyr; \citealt{2020AJ....160..108B}), recent observations have started to provide information on young super-Earth systems. \citet{2024AJ....168..239D} analyzed orbital period ratios using systems discovered by {\it TESS}, focusing on their age dependence. They showed that in systems older than 1 Gyr, only about 15\% of planet pairs lie near first-order MMRs, whereas in young systems ($<100$ Myr), about 70\% of pairs are close to first-order resonances. This result suggests that resonant configurations of super-Earths may be disrupted on a timescale of the order of 100 Myr. Although the number of observed young systems is still small, this information is valuable for developing theoretical models of super-Earth formation and subsequent orbital evolution. In this work, we investigate whether the observed time evolution of period ratios can be explained by $N$-body simulations, and if so, under what conditions.

In addition, we also focus on another important issue. As mentioned above, previous simulations have suggested that the lack of resonance in old super-Earth systems can be explained if resonant chains undergo dynamical instabilities. However, it is still unclear what physical processes trigger such instabilities. 
Several mechanisms have been proposed, including overstability of resonances caused by eccentricity damping \citep{2014AJ....147...32G, 2017MNRAS.468.3223X}, a large number of planets captured in the chain \citep{2018A&A...615A..63O}, perturbations from outer giant planets \citep{2021MNRAS.508..597P, 2024A&A...692A.246B}, the presence of higher-order resonances within the chain \citep{2025ApJ...982..111C}, changes in orbital architecture due to magnetospheric rebound \citep{2025arXiv250907866P}, and an increase in libration amplitudes for some reasons \citep{2025AJ....169..323L}. Each mechanism can disrupt resonant chains on different timescales and with different efficiencies. In this study, we also examine what triggers instabilities of resonant chains in our super-Earth formation simulations, and under what conditions such instabilities occur. We particularly focus on the influence of outer embryos on the inner super-Earths, which has also been investigated independently around the same time \citep{2025A&A_submitted_Goldberg}.

The structure of this paper is as follows. 
In Section~\ref{sec:model} we describe the numerical model used in our $N$-body simulations of super-Earth formation. 
In Section~\ref{sec:results} we present the simulation results. 
In Section~\ref{sec:secular} we investigate secular perturbations as a mechanism for triggering orbital instabilities, using both analytical expressions and additional $N$-body integrations, and discuss the conditions under which instabilities arise. 
Finally, in Section~\ref{sec:conclusions} we summarize our conclusions.

\section{Simulation model}\label{sec:model}

\subsection{Initial condition and orbital integration}
We consider a super-Earth formation model starting from a narrow ring of embryos \citep{2023NatAs...7..330B,2024ApJ...972..181O}, because such ring-origin models can explain various characteristics of planetary systems \citep[e.g.,][]{2022NatAs...6..357I,2023Icar..39615497W,2025ApJ...983...56G}. As the initial condition for the simulation, we follow \citet{2024ApJ...972..181O} and consider embryos that are randomly distributed in a ring between semimajor axes of $a=1$ and $1.5\,\mathrm{au}$. The total mass is set to $20\,M_\oplus$ (see Section~\ref{sec:n-body} for its impact), and we consider three cases in which the initial embryo mass is $M_{\rm ini} = 0.05$, $0.1$, and $0.2\,M_\oplus$. The central star has a mass of $1\,M_\odot$. For each condition, we perform ten runs with different initial positions of embryos. The initial eccentricities and orbital inclinations are small, about $0.01$.

The dynamical evolution of these embryos is investigated by $N$-body simulations that calculate mutual gravitational interactions. The basic equations of motion are given as follows:
\begin{eqnarray}
\frac{d^{2}\mathbf{r}_{i}}{dt^{2}} &=&
 - \frac{GM_{\ast}\mathbf{r}_{i}}{|\mathbf{r}_{i}|^{3}}
 - \sum_{j \ne i} \frac{GM_{j}(\mathbf{r}_{i} - \mathbf{r}_{j})}{|\mathbf{r}_{i} - \mathbf{r}_{j}|^{3}}
 - \sum_{j} \frac{GM_{j}\mathbf{r}_{j}}{|\mathbf{r}_{j}|^{3}}\nonumber \\
 &&+ \mathbf{F}_{\mathrm{mig}}
 + \mathbf{F}_{\mathrm{damp}},
\end{eqnarray}
where $\mathbf{r}_{i}$ is the position vector of the $i$-th body with respect to the central star, $M_{\ast}$ is the mass of the central star, $M_{j}$ is the mass of the $j$-th body, and $G$ is the gravitational constant.
The first term on the right-hand side represents the gravity of the central star, the second term represents the mutual gravity, and the third term represents the indirect term, which corrects for the acceleration of the central star induced by the planets’ gravity. In addition to these gravitational forces, we include forces due to the interaction between planets and the protoplanetary disk. The term $\mathbf{F}_{\mathrm{mig}}$ represents the additional force due to type I migration, and $\mathbf{F}_{\mathrm{damp}}$ denotes the additional force representing eccentricity and inclination damping by the gas disk. For type I migration, we adopt the formula of \citet{2011MNRAS.410..293P}, which includes the effect of the saturation of the corotation torque. For eccentricity and inclination damping, we follow the prescription of \citet{2008A&A...482..677C}. A more detailed introduction is provided in \citet{2015A&A...579A..65O}.

The above equations are solved using the fourth-order Hermite scheme with hierarchical time stepping \citep{1991ApJ...369..200M,1992PASJ...44..141M}. We also employ the Phantom GRAPE scheme \citep{2006NewA...12..169N} to accelerate the calculation of mutual gravitational interactions. When the distance between two bodies becomes smaller than the sum of their physical radii, we regard this as a collision and assume perfect accretion. We follow both planetary growth in the gas disk and orbital evolution after disk dispersal up to $100\,\mathrm{Myr}$.

\subsection{Disk model}
During the last ten years, it has become clear that planet formation strongly depends on the disk evolution model adopted. In particular, the orbital migration speed changes depending on the slope of the gas surface density, and planetary systems with different properties can form when the migration speeds are different \citep[e.g.,][]{2024ApJ...972..181O,2025AJ....170..180N}.

One of the suitable disk models for the formation of super-Earths is the disk that evolves under the effect of magnetically driven disk winds \citep{2016A&A...596A..74S,2018A&A...615A..63O}. To consider this disk model, we solve the following one-dimensional advection-diffusion equation in this study \citep{2016A&A...596A..74S,2020MNRAS.492.3849K}:
\begin{eqnarray}
\frac{\partial \Sigma_{\mathrm{g}}}{\partial t}
&=& \frac{1}{r} \frac{\partial}{\partial r} \left\{ 
\frac{2}{r \Omega} \left[
\frac{\partial}{\partial r} \left( r^{2} \Sigma_{\mathrm{g}} \overline{\alpha_{r\phi}} c_{s}^{2} \right)
+ r^{2} \overline{\alpha_{\phi z}} \frac{\Sigma_{\mathrm{g}} H \Omega^{2}}{2 \sqrt{\pi}}
\right] \right\}\nonumber \\
&&- \dot{\Sigma}_{\mathrm{MDW}}
- \dot{\Sigma}_{\mathrm{PEW}},
\end{eqnarray}
where $\Sigma_{\mathrm{g}}$ is the gas surface density, $r$ is the orbital radius, $\Omega$ is the angular velocity, $c_{s}$ is the sound speed, and $H$ is the disk scale height. 
The first term on the right-hand side represents viscous accretion, which is controlled by the parameter $\overline{\alpha_{r\phi}}$. The second term represents wind-driven accretion, which is determined by the parameter $\overline{\alpha_{\phi z}}$. The third and fourth terms represent mass loss due to magnetically driven disk winds and photoevaporative winds, respectively. The setup of each term follows \citet{2024ApJ...972..181O}, and for the viscous parameter $\overline{\alpha_{r\phi}}$, we adopt a value of $10^{-4}$.

As mentioned above, the result of super-Earth formation depends on the adopted disk model. In order to see the differences when different disk models are used, we also employ another model. A commonly used model is a power-law disk in which the surface density follows a simple power-law distribution. A recently used version of such a model is the two-component power-law disk model \citep[e.g.,][]{2015A&A...575A..28B,2019A&A...632A...7L}. In this model, the disk is divided into an inner region dominated by viscous heating and an outer region dominated by stellar irradiation, and each region is expressed with a different power-law distribution:
\begin{eqnarray}
\Sigma_{\mathrm{g,vis}} &=& 1320
\left( \frac{\dot{M}_{\mathrm{g}}}{10^{-9}\,M_{\odot}\,\mathrm{yr}^{-1}} \right)^{1/2}
\left( \frac{\overline{\alpha_{r\phi}}}{10^{-4}} \right)^{-3/4}\nonumber \\
&&\times \left( \frac{r}{1\,\mathrm{au}} \right)^{-3/8}
\ \mathrm{g\,cm^{-2}}, \\
\Sigma_{\mathrm{g,irr}} &=& 2500
\left( \frac{\dot{M}_{\mathrm{g}}}{10^{-9}\,M_{\odot}\,\mathrm{yr}^{-1}} \right)^{-1}
\left( \frac{\overline{\alpha_{r\phi}}}{10^{-4}} \right)^{-1}\nonumber \\
&&\times \left( \frac{r}{1\,\mathrm{au}} \right)^{-15/14}
\ \mathrm{g\,cm^{-2}},
\end{eqnarray}
where $\Sigma_{\mathrm{g,vis}}$ is the gas surface density in the viscous heating dominated region, $\Sigma_{\mathrm{g,irr}}$ is the gas surface density in the stellar irradiation dominated region, and $\overline{\alpha_{r\phi}} (=10^{-4})$ is the viscosity parameter. The gas accretion rate onto the central star is set to $\dot{M}_{\mathrm{g}} =10^{-9} \exp[-t/(0.5\,\mathrm{Myr})] M_\odot \,\mathrm{yr}^{-1}$, which mimics the dissipation of gas in the inner region (within 1 au) after approximately $t = 3\,\mathrm{Myr}$. For further detail, please refer to \citet{2023PSJ.....4...32O}.

\begin{figure}[ht!]
\plotone{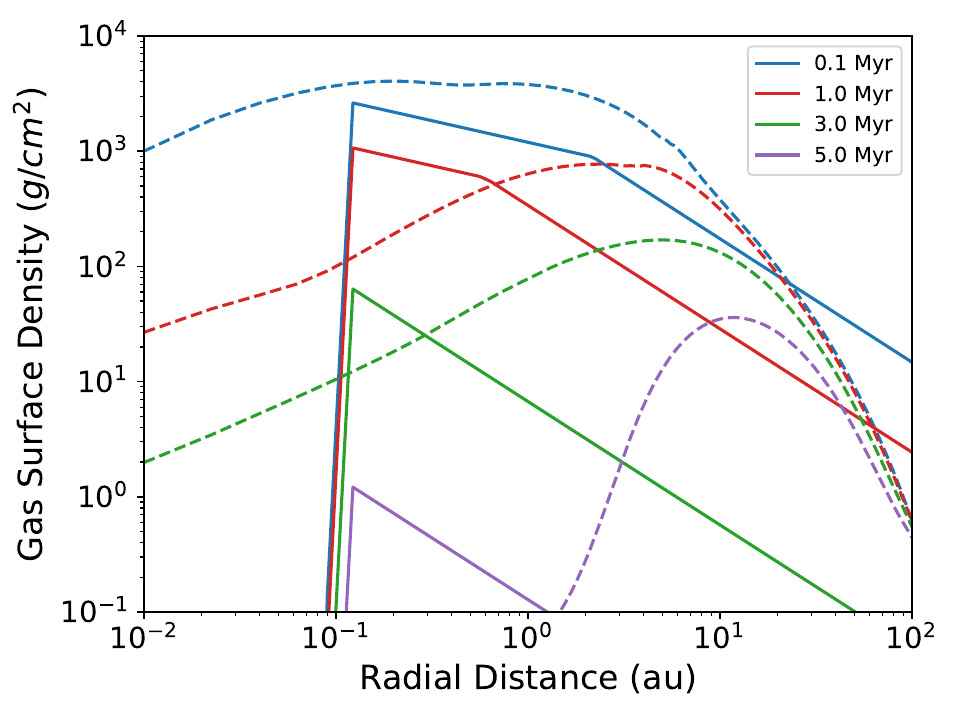}
\caption{Gas surface density profiles and their temporal evolution for two disk models. The solid line represents the two-component power-law model with an inner edge at $r = 0.1\,\mathrm{au}$, referred to as the rapid migration case. The dashed line represents the model evolving under the influence of disk winds, referred to as the slow migration case.}
\label{fig:disk}
\end{figure}

Figure~\ref{fig:disk} shows the time evolution of each disk model. In the case of a power-law distribution model, the slope of the gas surface density is $-3/8$ or $-15/14$, and in this case, planets with Earth mass undergo rapid inward migration on a timescale of about $0.1\,\mathrm{Myr}$ at $t=0.1\,{\rm Myr}$. In this disk, the inner edge is set at $r = 0.1\,\mathrm{au}$ using a hyperbolic tangent function with a width of $0.002\,\mathrm{au}$. On the other hand, in a disk evolving under the effect of disk winds, the slope of the gas surface density becomes flatter or even positive in the inner region, and inward migration is suppressed. In this paper, we refer to the case of a power-law distribution as the fast migration case, and the case of a disk evolving under the influence of disk winds as the slow migration case.

\section{Simulation results}\label{sec:results}
\subsection{Fast migration case}
\begin{figure*}[ht!]
\plotone{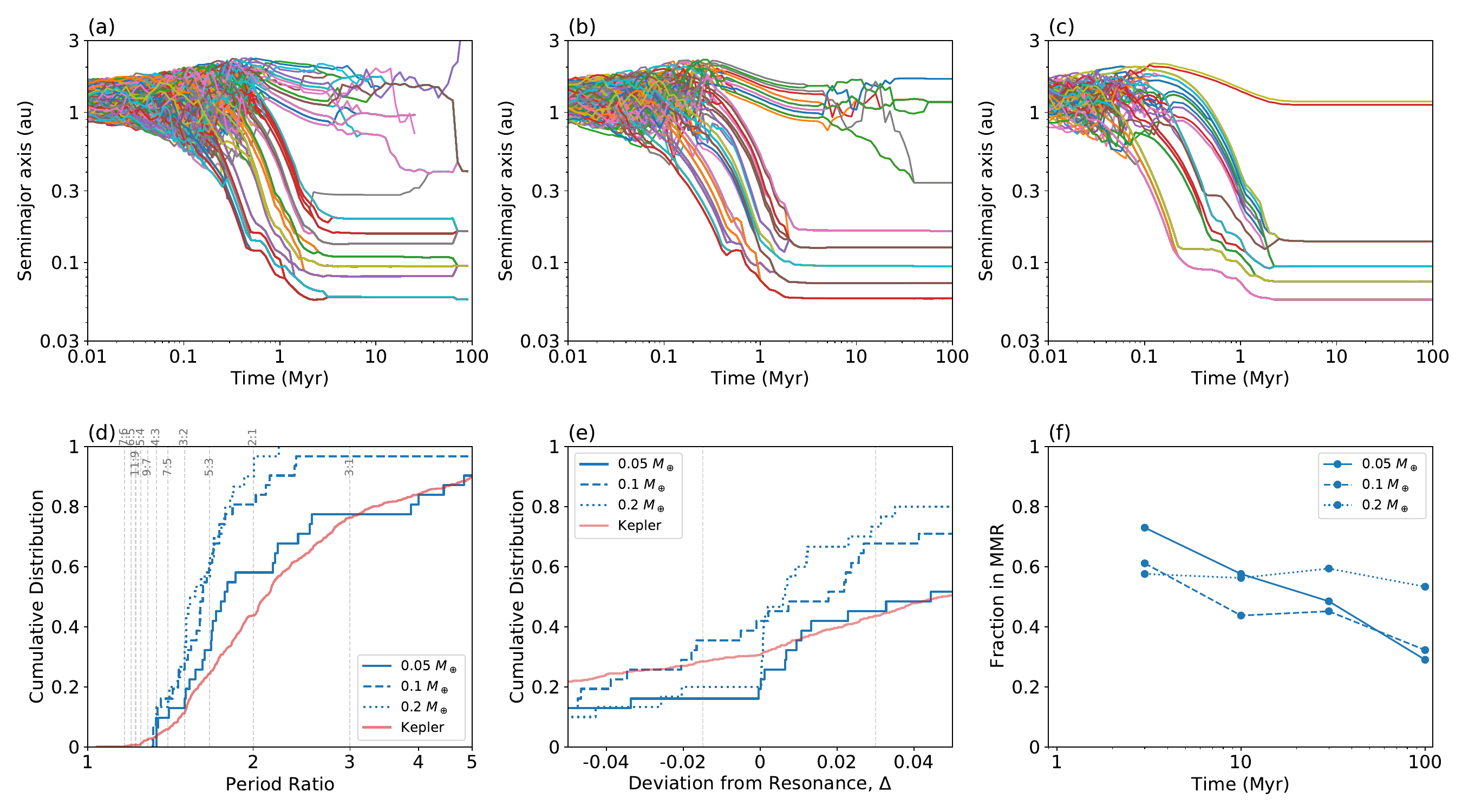}
\caption{Simulation results for the fast migration case. (a)–(c) Time evolution of the semimajor axes. Panels (a), (b), and (c) correspond to simulations with initial embryo masses of (a) $0.05\,M_{\oplus}$, (b) $0.1\,M_{\oplus}$, and (c) $0.2\,M_{\oplus}$, respectively. (d)–(f) Summary of ten simulation runs for each setup. (d) Cumulative distribution of orbital period ratios of super-Earth pairs at $t = 100\,\mathrm{Myr}$. Vertical lines indicate the locations of mean-motion resonances, and the red line shows the distribution from \textit{Kepler} observations. (e) Cumulative distribution of the offset $\Delta$ from the nearest integer period ratio for super-Earth pairs at $t = 100\,\mathrm{Myr}$. Vertical lines mark $\Delta = -0.015$ and $\Delta = 0.03$. (f) Fraction of super-Earth pairs in mean-motion resonances at $t = 3$, 10, 30, and $100\,\mathrm{Myr}$.}
\label{fig:100Myr_fast}
\end{figure*}

First, we examine the case in which planets undergo rapid inward migration in a disk modeled with a power-law surface density profile. Figure~\ref{fig:100Myr_fast}(a)–(c) show typical orbital evolution for initial embryo masses of $M_{\rm ini}= 0.05\,M_\oplus$, $0.1\,M_\oplus$, and $0.2\,M_\oplus$, respectively.
We begin by looking at Figure~\ref{fig:100Myr_fast}(a), where the initial embryo mass is $M_{\rm ini}= 0.05\,M_\oplus$. In this case, within the initial ring located at $a=1–2\,{\rm au}$, planets grow through giant impacts and reach around $1\,M_\oplus$ by $t \sim 0.1\,{\rm Myr}$. These planets then migrate inward due to type I migration. The first planet to migrate inward reaches the inner edge of the disk around $t \simeq 0.5\,{\rm Myr}$ and stops just outside $a=0.1\,{\rm au}$ because the planet experiences a positive corotation torque from the inner edge of the disk \citep[][]{2021A&A...648A..69A}. This planet captures the next inward-migrating planet into a mean-motion resonance and forms a resonant chain. The chain may grow longer by capturing other late-arriving planets. If capture into resonance fails, the newly arrived planet collides with the planets in the chain, which leads to rearrangement of the resonant system. Through such rearrangement processes or simply by being pushed by outer migrating planets, the resonant chain can move slightly inward, even inside the disk’s inner edge. By the time the disk disperses around $t \simeq 3\,{\rm Myr}$, the inward migration of planets formed in the ring has stopped. At this stage, many of the resonant planet pairs are found in first-order commensurabilities, typically between 3:2 and 5:4, although second-order resonances also occur in some cases.
This phase of inward migration within the disk occurs similarly in Figure~\ref{fig:100Myr_fast}(b) and (c), where the initial embryo mass differs. The main difference is the amount of remnant embryos left in the outer region. When the initial embryo mass is smaller, super-Earths tend to scatter smaller embryos outward during the growth phase from the ring. As a result, more embryos remain around $a \sim 1\,{\rm au}$ when the initial embryo's mass is small. We will discuss this point in more detail later.

After the disk disperses ($t>3\,{\rm Myr}$), the long-term stability over about $100\,{\rm Myr}$ is also important. In the case of Figure~\ref{fig:100Myr_fast}(a), dynamical instability occurs in the resonant chain of super-Earths around $t=60–70\,{\rm Myr}$. This leads to orbit crossing and collisions between super-Earths. As a result, four planets with more than one Earth mass are finally formed inside $a=0.5\,{\rm au}$, and they are no longer in mean-motion resonances. On the other hand, in Figures~\ref{fig:100Myr_fast}(b) and (c), no dynamical instability occurs during the long-term evolution after disk dispersal. In these cases, 4--5 super-Earths are formed near period ratios close to 3:2 and 5:3.

Figures~\ref{fig:100Myr_fast}(d)--(f) summarize the period ratios of super-Earths ($a<1\,{\rm au},\,M>1\,M_\oplus$) in the case of fast migration. For each initial mass, we performed 10 runs with different initial positions of the embryos, and the results are shown as cumulative distributions. Figure~\ref{fig:100Myr_fast}(d) shows the cumulative distribution of period ratios between adjacent super-Earths at $t=100\,{\rm Myr}$. In all cases, the systems are more compact than the observed distribution of super-Earths, with smaller period ratios. In particular, when the initial mass is $0.1\,M_\oplus$ or $0.2\,M_\oplus$, the resonant chains are rarely broken during long-term evolution, and the systems remain much more compact compared with observations.

Figure~\ref{fig:100Myr_fast}(e) shows how far the period ratios of adjacent super-Earths are from the nearest integer ratio. Here we define planets to be in resonance when $\Delta [\equiv (P_{\rm out}/P_{\rm in})/(\mathrm{integer\ ratio}) - 1]$ is between $-0.015$ and $0.03$ \citep{2024AJ....168..239D}. At $t=100\,{\rm Myr}$, the fraction of super-Earth pairs in resonance is 29\%, 32\%, and 53\% for the cases with initial masses of $0.05\,M_\oplus$, $0.1\,M_\oplus$, and $0.2\,M_\oplus$, respectively. This is larger than the observed fraction of resonant pairs, which is about 15\%. From the shape of the cumulative distribution, we also see a difference from the observed distribution, especially in the case of $0.2\,M_\oplus$. 
Note that a pile-up just outside resonance (seen as a change of slope in the cumulative distribution around $\Delta=0$) is also visible in the cases with $0.05\,M_\oplus$ and $0.2\,M_\oplus$. One proposed explanation for the origin of the pile-up just exterior to exact orbital period ratios is the ping-pong mechanism involving Mercury-size bodies \citep{2024ApJ...971....5W}. In our study, it is also possible that this pile-up is triggered by interactions between high-eccentricity embryos and super-Earths. This point will be examined in a separate paper.

Figure~\ref{fig:100Myr_fast}(f) shows the time evolution of the fraction of super-Earth pairs in resonance, as determined from $\Delta$. As expected from Figure~\ref{fig:100Myr_fast}(a)–(c), the fraction of resonant pairs decreases with time. The largest change is in the case of $0.05\,M_\oplus$: the fraction is 73\% at $t=3\,{\rm Myr}$ but decreases to 29\% at $t=100\,{\rm Myr}$. In contrast, when the initial mass is $0.1\,M_\oplus$ or $0.2\,M_\oplus$, the decrease is small. Especially for $0.2\,M_\oplus$, the fraction of resonant pairs remains almost the same between $t=3\,{\rm Myr}$ and $t=100\,{\rm Myr}$. 
In summary, for the case of fast migration shown in Figure~\ref{fig:100Myr_fast}, when the initial mass is $0.05\,M_\oplus$, the system shows a trend closer to observations because dynamical instability occurs after disk dispersal. However, in all cases, the results are not fully consistent with the observed period-ratio distribution of super-Earths.

\subsection{Slow migration case}
\begin{figure*}[ht!]
\plotone{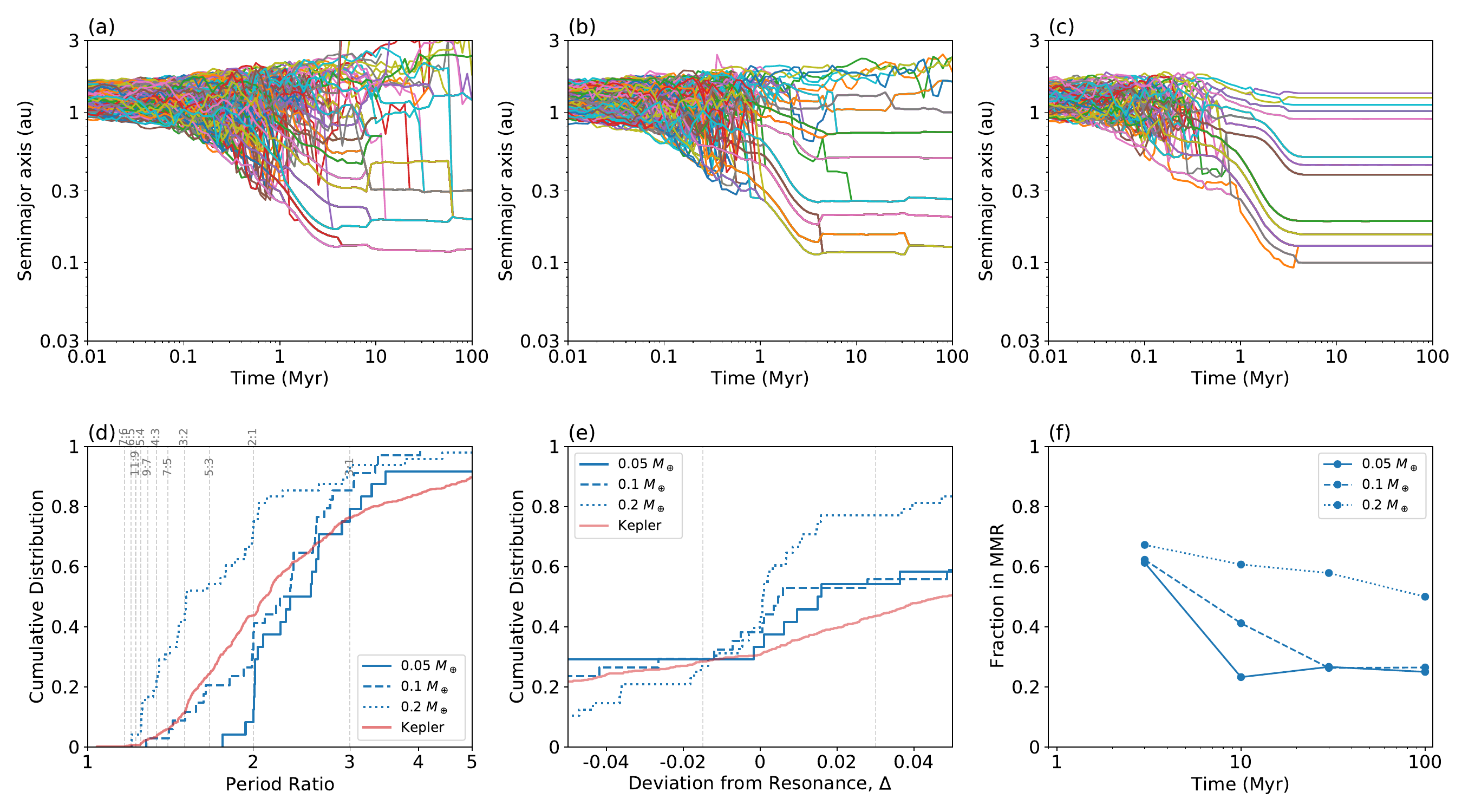}
\caption{Same as Figure~\ref{fig:100Myr_fast}, but showing the results for the slow migration case.}
\label{fig:100Myr_slow}
\end{figure*}

Next, we show results for slow migration in a wind-evolving disk. Figure~\ref{fig:100Myr_slow}(a)–(c) show typical outcomes for simulations starting with different initial embryo masses. The qualitative results are not very different from those in the fast migration case shown in Figure~\ref{fig:100Myr_fast}. 
During the disk phase ($t<3\,{\rm Myr}$), planets grow at the location of the initial ring and start inward migration once they reach about $1\,M_\oplus$. Because the gas surface density gradient is shallower, the migration speed is slower than in Figure~\ref{fig:100Myr_fast}. In this disk model, we do not assume a sharp inner edge. However, since super-Earths migrate at different speeds, they can still be captured into mean-motion resonances with neighboring planets. At $t=3\,{\rm Myr}$, just before disk dispersal, many super-Earth pairs are in resonant chains (e.g., 3:2, 4:3, 5:4, 6:5).

A major difference from the fast migration case is that dynamical instability often occurs between $t=3-100\,{\rm Myr}$ after disk dispersal. For example, in the case with $M_{\rm ini}=0.05\,M_\oplus$ shown in Figure~\ref{fig:100Myr_slow}(a), collisions between super-Earths occur around $t=10\,{\rm Myr}$ and $70\,{\rm Myr}$. As a result, three super-Earths are finally formed, which are no longer in resonances. In the case with $M_{\rm ini}=0.1\,M_\oplus$ (Figure~\ref{fig:100Myr_slow}(b)), dynamical instability also occurs at around $t=30\,{\rm Myr}$. On the other hand, for $M_{\rm ini}=0.2\,M_\oplus$ (Figure~\ref{fig:100Myr_slow}(c)), the resonant chain remains stable for a long time, and the final system still has period ratios close to resonance.

Figure~\ref{fig:100Myr_slow}(d)--(f) summarize the results of 10 runs for each initial mass in the case of slow migration, shown as cumulative distributions. Figure~\ref{fig:100Myr_slow}(d) shows that, compared with the fast migration case, the final period ratios of adjacent super-Earths tend to be larger. This is especially true for initial masses of $0.05\,M_\oplus$ and $0.1\,M_\oplus$, where dynamical instability increases the period ratios and makes the distribution closer to the observed one. 
Figure~\ref{fig:100Myr_slow}(e) shows the deviation from exact integer ratios. The final fraction of resonant pairs is 25\%, 26\%, and 50\% for $M_{\rm ini}=0.05\,M_\oplus$, $0.1\,M_\oplus$, and $0.2\,M_\oplus$, respectively, which is smaller than in the fast migration case. 
A pile-up is also seen just outside exact resonance, where the slope of the cumulative distribution changes at positive $\Delta$. In particular, for $0.05\,M_\oplus$ and $0.1\,M_\oplus$, the shape of the cumulative distribution is more consistent with observations. 
Figure~\ref{fig:100Myr_slow}(f) shows the time evolution of the resonant fraction. For $0.05\,M_\oplus$ and $0.1\,M_\oplus$, the fraction clearly decreases with time: about 60\% just after formation, but less than 30\% by $t=100\,{\rm Myr}$. This is similar to the observed trend suggested by \citet{2024AJ....168..239D}.

From these results, we find that super-Earths formed from the ring are captured into mean-motion resonances during inward migration by type I migration. In some cases, dynamical instability occurs after disk dispersal. When this happens, the final distribution of period ratios, as well as its time evolution, shows a trend consistent with the observed distribution. Overall, looking at Figures~\ref{fig:100Myr_fast} and \ref{fig:100Myr_slow}, the closest agreement with observations is obtained in the case of slow migration with $M_{\rm ini}=0.05\,M_\oplus$ or $0.1\,M_\oplus$, followed by fast migration with $M_{\rm ini}=0.05\,M_\oplus$. In other cases, no clear dynamical instability is seen in most simulations, and the resulting distributions of period ratios are different from observations.

\subsection{Factors leading to dynamical instability}\label{sec:factors}
\begin{figure*}[ht!]
\epsscale{0.8}
\plotone{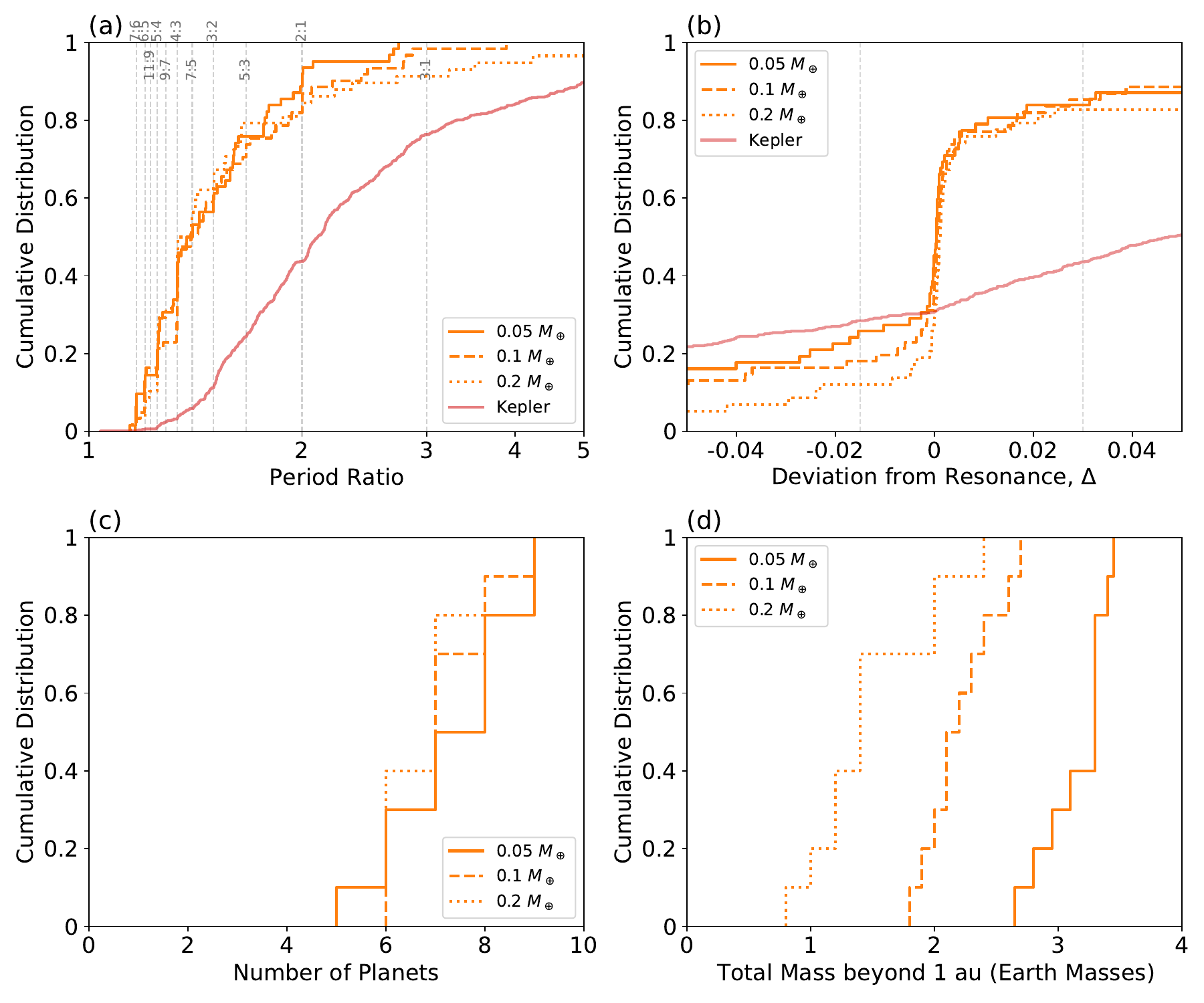}
\caption{Properties of the systems at $t = 3\,\mathrm{Myr}$. (a) Cumulative distribution of the orbital period ratios of adjacent super-Earth pairs. (b) Cumulative distribution of the deviations from exact commensurabilities for adjacent super-Earth pairs. (c) Cumulative distribution of the number of super-Earths per system. (d) Cumulative distribution of the total mass of embryos that remain on outer orbits with $a > 1\,\mathrm{au}$.}
\label{fig:3Myr_slow}
\end{figure*}

So far, we have found that the observed features can be explained when mean-motion resonances are broken by dynamical instability after disk dispersal. Then, under what conditions does dynamical instability occur? In this subsection, we focus on the case of slow migration and examine the properties of systems at $t=3\,{\rm Myr}$, just before disk dispersal. The results for fast migration are shown in Appendix~\ref{sec:appA}.

Figure~\ref{fig:3Myr_slow} shows the cumulative distributions of system properties at $t=3\,{\rm Myr}$ in the slow migration case. Figure~\ref{fig:3Myr_slow}(a) shows the distribution of period ratios of adjacent super-Earths. Previous studies of orbital stability have shown that systems with smaller period ratios (i.e., more closely spaced orbits) tend to become unstable on shorter timescales, both in non-resonant systems \citep{1996Icar..119..261C} and in resonant systems \citep{2012Icar..221..624M}. Thus, one might expect that in the cases with dynamical instability (e.g., $M_{\rm ini}=0.05\,M_\oplus$ or $0.1\,M_\oplus$), the period ratios would be smaller. However, Figure~\ref{fig:3Myr_slow}(a) shows that there is no significant difference in period ratios among the three initial masses at $t=3\,{\rm Myr}$. Therefore, the difference between stable and unstable outcomes in our simulations cannot be explained by period ratios or orbital separations.

Next, Figure~\ref{fig:3Myr_slow}(b) shows the deviation from exact commensurabilities. Previous dynamical simulations have shown that the further a planet pair is from exact commensurability, the shorter the timescale for resonance to become unstable \citep{2020ApJ...893...43M}. It has also been suggested that systems with a smaller fraction of resonant pairs tend to be more unstable \citep{2025ApJ...982..111C}. However, Figure~\ref{fig:3Myr_slow}(b) shows that the fraction of super-Earth pairs in resonance (defined as $-0.015 < \Delta < 0.03$) does not change much among the three cases. Although the case with $M_{\rm ini}=0.2\,M_\oplus$ has about 10\% more resonant pairs, the difference is small.

Figure~\ref{fig:3Myr_slow}(c) shows the number of super-Earths. It is known that the number of planets strongly affects dynamical stability \citep{1996Icar..119..261C, 2012Icar..221..624M}. However, in our results, there is no clear difference in the number of super-Earths at $t=3\,{\rm Myr}$ among the three initial masses. Thus, the difference in outcomes seen in Figure~\ref{fig:100Myr_slow} also cannot be explained by the number of planets.

We find that the most significant difference at the time of disk dispersal is the total mass of embryos remaining in the outer region. Figure~\ref{fig:3Myr_slow}(d) shows the total mass of bodies beyond $a=1\,{\rm au}$. In the case with $M_{\rm ini}=0.05\,M_\oplus$, which often leads to dynamical instability after the disk dispersal, many embryos remain in the outer orbits. As we will show in the next section, these outer embryos excite the eccentricities of inner super-Earths through secular perturbations, which eventually trigger dynamical instability. Although the possibility that outer giant planets can induce dynamical instability among inner super-Earths has been discussed, we find that high-eccentricity embryos can also cause such instabilities\footnote{The dynamical instability of inner resonant chains induced by outer small planets has also been investigated independently around the same time by \citet{2025A&A_submitted_Goldberg}.}. These outer embryos are produced when small embryos are scattered outward during the growth of super-Earths from the ring near $a=$1\,au. The smaller the embryo mass (i.e., the larger the mass ratio between super-Earths and embryos), the more easily they are scattered outward \citep[][]{2020ApJ...899...91O, 2025ApJ_submitted_He}. This explains why Figure~\ref{fig:3Myr_slow}(d) shows that smaller initial masses leave more embryos in the outer region.

\section{Secular perturbations}\label{sec:secular}

\subsection{Analytical framework for secular interactions}\label{sec:ana}
We have found that remaining outer embryos can influence the long-term stability of inner super-Earths. This effect is likely explained by secular perturbations. Therefore, in this section, we outline the basics of secular resonance theory, mainly based on the Laplace–Lagrange theory \citep[e.g.,][]{1961mcm..book.....B,1999ssd..book.....M}.

In a system of $N$ particles, the eccentricity of particle $i$ at a given time $t$ is expressed as
\begin{equation}\label{eq:e}
    e_i(t) = \left[ h_i(t)^2 + k_i(t)^2 \right]^{1/2},
\end{equation}
where the components of the eccentricity vector are defined using the longitude of pericenter, $\varpi_i(t)$, as
\begin{equation}
h_i(t) = e_i(t)\,\sin \varpi_i(t), \quad k_i(t) = e_i(t)\,\cos \varpi_i(t).
\end{equation}
The secular part of the disturbing function for particle $i$, retaining only the second-order terms in eccentricity, can be written as
\begin{equation}\label{eq:disturbing}
R_i = n_i a_i^2 \left[ \frac{1}{2} A_{ii} (h_i^2 + k_i^2) + \sum_{j=1,\,j \neq i}^N A_{ij}
(h_i h_j + k_i k_j)\right].
\end{equation}
The coefficients are given by
\begin{eqnarray}
A_{ii} &=& \frac{n_i}{4} \sum_{j=1,\,j \neq i}^N \frac{M_j}{M_\ast + M_i} \alpha_{ij} \bar{\alpha}_{ij} b_{3/2}^{(1)}(\alpha_{ij}), \\
A_{ij} &=& - \frac{n_i}{4} \frac{M_j}{M_\ast + M_i} \alpha_{ij} \bar{\alpha}_{ij} b_{3/2}^{(2)}(\alpha_{ij}),
\end{eqnarray}
where
\begin{equation}
  \alpha_{ij} =
  \left\{
    \begin{array}{ll}
      a_j / a_i & \text{if } a_i > a_j, \\
      a_i / a_j & \text{if } a_i < a_j,
    \end{array}
  \right.
\end{equation}
\begin{equation}
  \bar{\alpha}_{ij} =
  \left\{
    \begin{array}{ll}
      1 & \text{if } a_i > a_j, \\
      a_i / a_j & \text{if } a_i < a_j.
    \end{array}
  \right.
\end{equation}
The Laplace coefficients are
\begin{eqnarray}
b_{3/2}^{(1)}(\alpha) &=& \frac{1}{\pi} \int_0^{2\pi} \frac{\cos\psi \, d\psi}{(1 - 2\alpha \cos \psi + \alpha^2)^{3/2}}, \\
b_{3/2}^{(2)}(\alpha) &=& \frac{1}{\pi} \int_0^{2\pi} \frac{\cos(2\psi) \, d\psi}{(1 - 2\alpha \cos \psi + \alpha^2)^{3/2}}.
\end{eqnarray}
In this case, the eccentricity vector can be expressed using the eigenvalues 
$g_1, g_2, \ldots$ of the $N \times N$ matrix $\mathbf{A}=(A_{ij})$. 
Let $\bar{\boldsymbol e}^{(j)} = (\bar e_{1j}, \bar e_{2j}, \ldots, \bar e_{Nj})$ 
denote the $j$-th eigenvector of $\mathbf{A}$. Then,
\begin{eqnarray}
     h_i(t) &=& \sum_{j=1}^N e_{ij}\,\sin(g_j t + \beta_j), \\
     k_i(t) &=& \sum_{j=1}^N e_{ij}\,\cos(g_j t + \beta_j),   
\end{eqnarray}
where the amplitude of each mode is given by
\begin{equation}
    e_{ij} = S_j\,\bar e_{ij},
\end{equation}
where the coefficients $S_j$ and phases $\beta_j$ are determined by the initial conditions. From the above, if the system is governed solely by secular interactions among the $N$ particles, the eccentricity at any given time can be calculated by evaluating Equation~(\ref{eq:e}) once the initial conditions are specified.

Next, we consider two planets orbiting a central star of mass $M_\ast$ (see Section~\ref{sec:n-body} for a discussion of the number of perturbers). The planets have masses $M_1$ and $M_2$, and semimajor axes satisfying $a_1 < a_2$. The mean motion of each planet is given by $n_i = \sqrt{G M_\ast / a_i^3}$. We focus on the case with initial conditions $e_1(0) = 0$ and $e_2(0) = e_{20} \ne 0$, and examine the amplitude and period of the eccentricity oscillation of the inner planet ($i = 1$). By expressing the eccentricity vector as a complex number, $z_i = e_i \exp(i \varpi_i)$, the time evolution of the inner planet ($i=1$) is
\begin{eqnarray}
z_1(t) &=& \sum_{j=1}^2 S_j\,\bar e_{1j}\,\exp(i g_j t),
\end{eqnarray}
where the mode coefficients $S_j$ are fixed by the initial conditions; 
in our normalization this yields 
\begin{eqnarray}
z_1(t) = e_{20} \sum_{j=1}^2 \bar e_{2j}\,\bar e_{1j}\,\exp(i g_j t).
\end{eqnarray}
Using this relation, the maximum eccentricity of particle $i$ (here, $e_{1,\max}$) can be calculated as
\begin{equation}
    e_{1,\max} = e_{20} \Bigl| \sum_{j=1}^2 \bar e_{2j}\,\bar e_{1j} \Bigr|.
\end{equation}
By using the expressions for the components of the eigenvectors,
\begin{eqnarray}
  \bar e_{1j} &=& \frac{A_{12}}{\sqrt{(g_j - A_{11})^2 + A_{12}A_{21}}}, \\
  \bar e_{2j} &=& \frac{g_j - A_{11}}{\sqrt{(g_j - A_{11})^2 + A_{12}A_{21}}},
\end{eqnarray}
the maximum eccentricity of the inner planet becomes
\begin{eqnarray}
  e_{1,\max}
  &=& e_{20} \sum_{j=1}^2
     \frac{|A_{12}\,(g_j - A_{11})|}{(g_j - A_{11})^2 + A_{12}A_{21}}\\
  &=& \frac{2\,|A_{12}|}{\sqrt{(A_{11} - A_{22})^2 + 4\,A_{12}A_{21}}}\, e_{20}\\
  &=& C_{12} \, e_{20},
  \label{eq:e1max}
\end{eqnarray}
where we define
\begin{equation}
   C_{12} \equiv \frac{2\,|A_{12}|}{\sqrt{(A_{11} - A_{22})^2 + 4\,A_{12}A_{21}}}.
\end{equation}
Using the matrix elements, we can compute $e_{1,\max}$ explicitly:
\begin{eqnarray}
\label{eq:A11}
A_{11} &=& \frac{n_1}{4}\frac{M_2}{M_\ast+M_1}\,\alpha^2\,b_{3/2}^{(1)}(\alpha),\\
A_{12} &=& -\frac{n_1}{4}\frac{M_2}{M_\ast+M_1}\,\alpha^2\,b_{3/2}^{(2)}(\alpha),\\
A_{21} &=& -\frac{n_2}{4}\frac{M_1}{M_\ast+M_2}\,\alpha\,b_{3/2}^{(2)}(\alpha),\\
A_{22} &=& \frac{n_2}{4}\frac{M_1}{M_\ast+M_2}\,\alpha\,b_{3/2}^{(1)}(\alpha),
\label{eq:A22}
\end{eqnarray}
where $\alpha=a_1/a_2$. If the outer planet is sufficiently massive, Equation~(\ref{eq:e1max}) can be simplified as
\begin{eqnarray}\label{eq:e1max2}
  e_{1,\max}
  \simeq \frac{2|A_{12}|}{|A_{11} - A_{22}|}\, e_{20} 
  \sim 2\,\frac{b_{3/2}^{(2)}(\alpha)}{b_{3/2}^{(1)}(\alpha)}\, e_{20}.
\end{eqnarray}

We now examine the timescale of the eccentricity oscillation in the case of two interacting particles. If a particular mode ($j_*$) dominates the amplitude for the inner planet, the observed secular period is approximately
\begin{equation}
  T_{\rm sec} = \frac{2\pi}{g_{j_*}}.
\end{equation}
If the outer planet is sufficiently massive and the $j = 2$ mode dominates, this can be further approximated as
\begin{equation}\label{eq:Tsec}
  T_{\rm sec} \sim \frac{2\pi}{A_{11}}
  \simeq  \frac{8\pi}{n_1}\,\frac{M_\ast}{M_2}\,\frac{1}{\alpha^2\,b_{3/2}^{(1)}(\alpha)}.
\end{equation}

\begin{figure*}[ht!]
\plotone{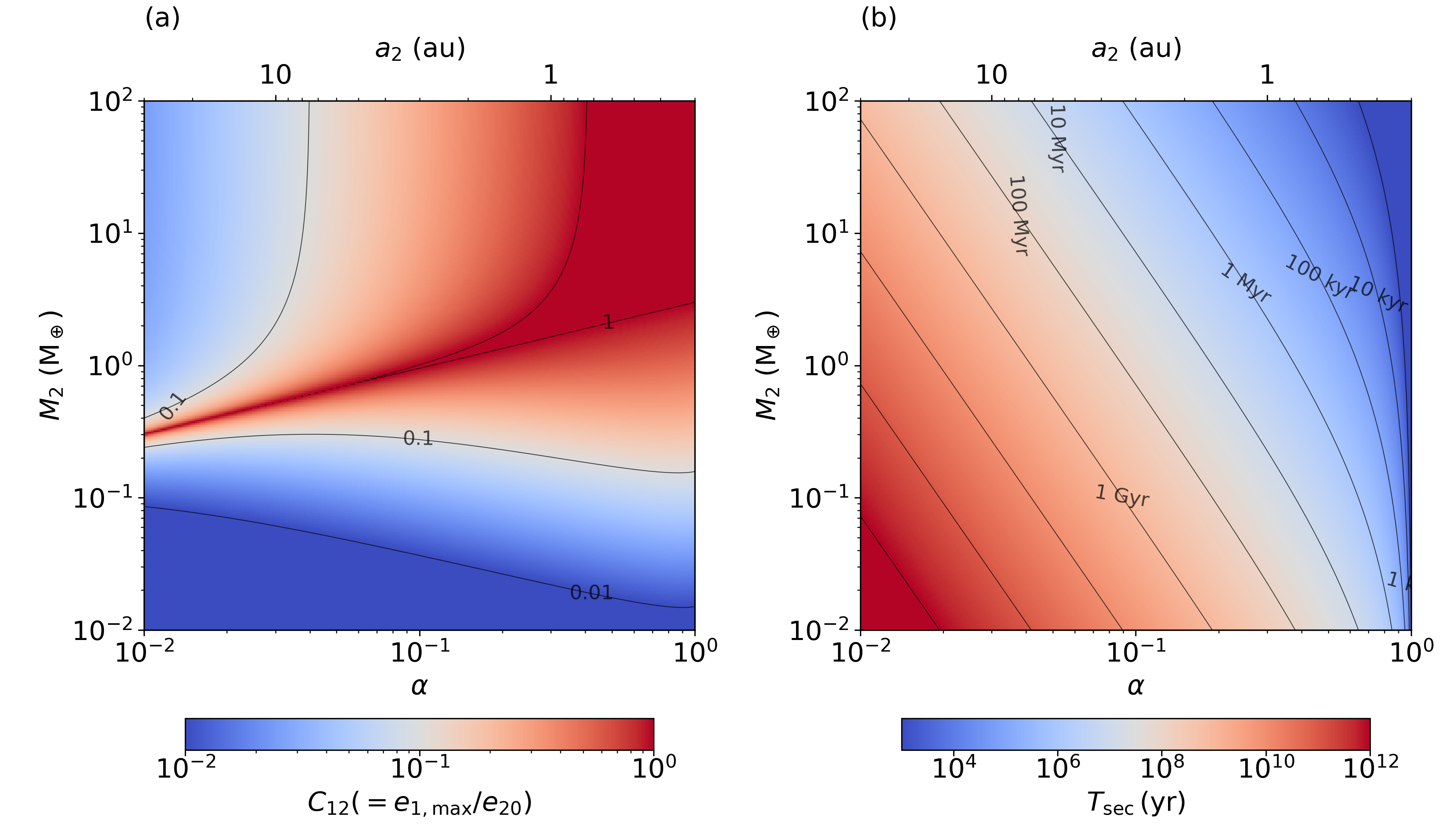}
\caption{(a) Plot of the coefficient $C_{12} (=e_{1,\max}/e_{20})$ from Equation~(\ref{eq:e1max}), evaluated using Equations~(\ref{eq:A11})–(\ref{eq:A22}). (b) Approximate secular timescale obtained from Equation~(\ref{eq:Tsec}), assuming $a_1 = 0.3\,\mathrm{au}$ and $M_1 = 3\,M_\oplus$.}
\label{fig:ana}
\end{figure*}

From these equations, we can predict how the amplitude and timescale of eccentricity oscillations of inner super-Earths depend on the properties of the outer planet. First, regarding the amplitude, as seen in Equations~(\ref{eq:e1max}) and (\ref{eq:e1max2}), it is proportional to the initial eccentricity of the outer planet $e_{20}$. Therefore, if $e_{20}$ is large, the maximum eccentricity of the inner planet $e_{1,\max}$ can also become large. Regarding the dependence on the mass of the outer planet $M_2$, when $M_2$ is much larger than $M_1$, Equation~(\ref{eq:e1max2}) shows that $e_{1,\max}$ does not depend on the mass. On the other hand, when $M_2$ is smaller than $M_1$, Equation~(\ref{eq:e1max}) indicates that $e_{1,\max}$ becomes smaller as $M_2$ decreases. Figure~\ref{fig:ana}(a) shows the coefficient from Equation~(\ref{eq:e1max}) for $M_1 = 3\,M_\oplus$ at $a_1 = 0.3\,\mathrm{au}$, plotted as a function of different values of $\alpha$ and $M_2$. One can clearly see that the mass dependence changes around $M_2 \sim 0.1\,M_\oplus$, a trend that is directly apparent in the figure. In addition, regarding the dependence on $\alpha$ (which corresponds to the semimajor axis of the outer planet $a_2$), when $M_2$ is much larger than $0.1\,M_\oplus$, Equation~(\ref{eq:e1max2}) shows that $e_{1,\max}$ decreases as $\alpha$ becomes smaller. On the other hand, when $M_2$ is smaller than $0.1\,M_\oplus$, Equation~(\ref{eq:e1max}) suggests that the dependence of $e_{1,\max}$ on $\alpha$ becomes weaker compared to Equation~(\ref{eq:e1max2}). This trend is also evident in Figure~\ref{fig:ana}(a).

Next, we examine the timescale of the eccentricity variation of the inner planet. According to Equation~(\ref{eq:Tsec}), the timescale does not depend on the eccentricity. The equation also shows that the timescale is inversely proportional to $M_2$ and becomes shorter as $M_2$ increases. In addition, the timescale strongly depends on $\alpha$. For example, if $\alpha$ becomes three times smaller (i.e., if $a_2$ becomes three times larger), the timescale is estimated to become roughly 30 times longer. Figure~\ref{fig:ana}(b) shows Equation~(\ref{eq:Tsec}) plotted for different values of $\alpha$ and $M_2$. From this figure, it is also evident that the timescale is a strong function of $\alpha$.

\subsection{\textit{N}-body simulation}\label{sec:n-body}

\begin{figure*}[ht!]
\plotone{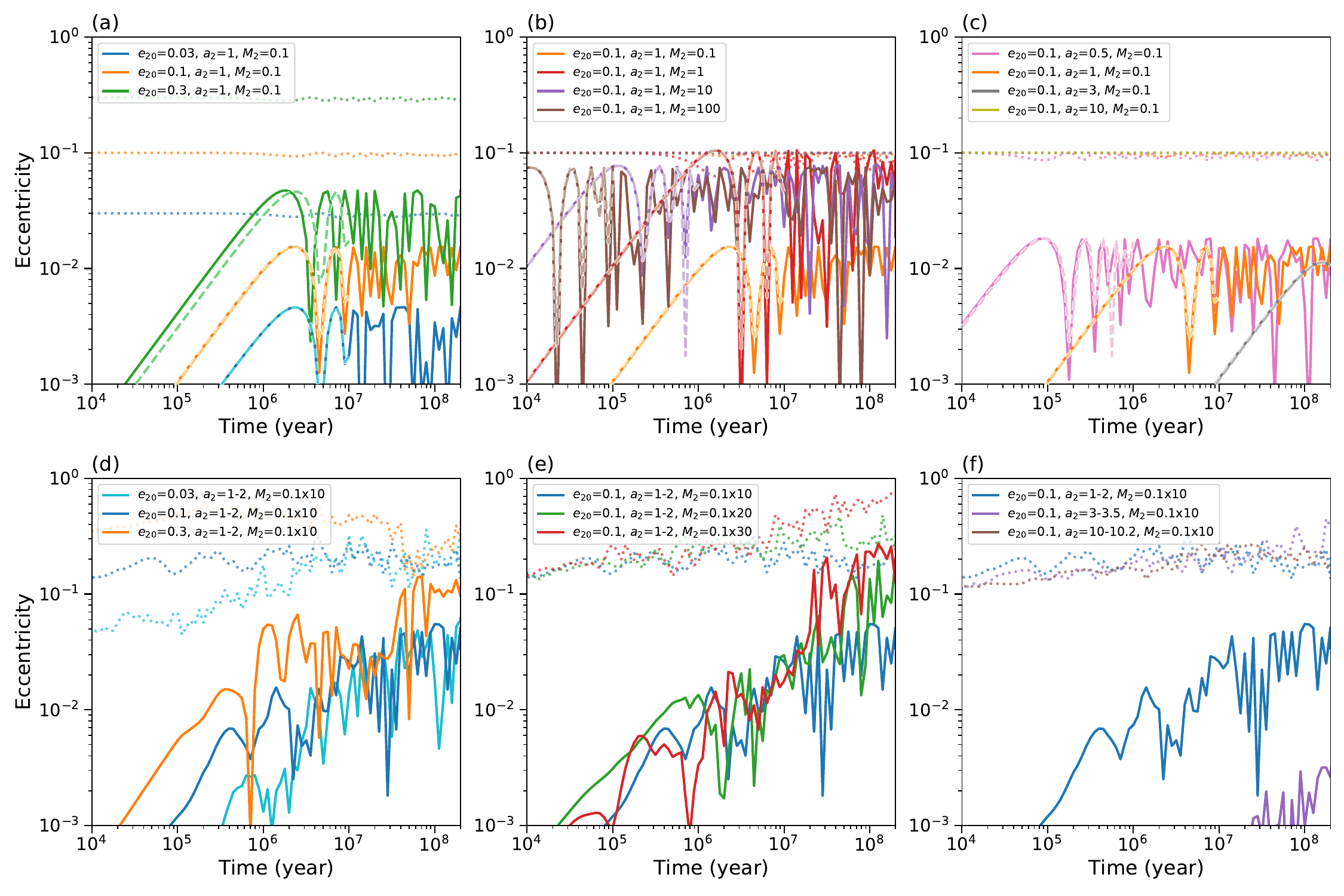}
\caption{(a)–(c) Time evolution of the eccentricity of an inner super-Earth with $a_1 = 0.3\,\mathrm{au}$ and $M_1 = 3\,M_\oplus$, perturbed by an outer planet with different values of $a_2$, $M_2$, and $e_{20}$. Solid lines show the simulated eccentricity of the inner super-Earth, dashed lines show the evolution predicted by the analytic expression, and dotted lines show the eccentricity of the outer perturber. The analytic solutions are shown only for the first few oscillation cycles in each case. (d)–(f) Results of $N$-body simulations with multiple outer perturbers. Solid lines indicate the eccentricity evolution of the inner super-Earth, while dotted lines indicate the maximum eccentricity of the outer perturbers. All panels are plotted with a logarithmic spacing of 0.05 dex intervals.}
\label{fig:secular}
\end{figure*}

We now investigate how the eccentricity of the inner super-Earth is affected by secular perturbations from planetary embryos located in outer orbits. As a first step, we consider the case of a single outer perturber and compare the results with the analytical predictions derived in Section~\ref{sec:ana}. Then, we extend the analysis to systems with multiple outer embryos. In such multiple-embryo systems, the eccentricities of the outer bodies change over time due to mutual gravitational scattering, so it becomes difficult to apply the analytical treatment discussed earlier.
Note that in the \textit{N}-body simulations presented in this section, the effects of the gas disk and its complicated evolution are not included. Therefore, we use the \texttt{REBOUND} code with the \texttt{MERCURIUS} integrator to follow the orbital evolution for up to 200\,Myr \citep{2012A&A...537A.128R,2019MNRAS.485.5490R}. The maximum timestep is set to 1/100 of the orbital period at $a=$ 0.3\,au. Unlike the simulations in Section~\ref{sec:results}, only one run is performed for each setting in this section.

Figure~\ref{fig:secular}(a) shows the evolution of the eccentricity of the inner super-Earth and the maximum eccentricity of the outer embryo when a single outer embryo with a mass of $0.1\,M_\oplus$ is placed at $a_2 = 1\,\mathrm{au}$, and the initial eccentricity $e_{20}$ is varied. In the case of $e_{20} = 0.1$, the eccentricity of the inner super-Earth exhibits oscillations on a timescale of 10 Myr. This value is comparable to that roughly estimated using Equation~(\ref{eq:Tsec}) (see also Figure~\ref{fig:ana}(b)). The eccentricity of the outer perturber also shows oscillatory behavior. As discussed in Section~\ref{sec:ana}, the oscillation timescale does not depend on $e_2$, and indeed it remains roughly the same for $e_{20} = 0.03$, 0.1, and 0.3. On the other hand, the amplitude of the oscillations increases with larger $e_{20}$. 
Note also that when $e_{20}$ is small, the numerical results match the analytical solution well. For higher values such as $e_{20} = 0.3$, however, the deviation becomes more noticeable. This discrepancy arises because the perturbation potential used in Equation~(\ref{eq:disturbing}) is truncated at second order in eccentricity.

Figure~\ref{fig:secular}(b) shows the results when the mass of the perturber is varied while keeping $a_2 = 1\,\mathrm{au}$ and $e_{20} = 0.1$. The amplitude of the eccentricity oscillations remains nearly the same, except for the case of $M_2 = 0.1\,M_\oplus$. This is consistent with the prediction from Section~\ref{sec:ana}: when $M_2 \lesssim 0.1\,M_\oplus$, the maximum eccentricity $e_{1,\max}$ depends on $M_2$, but otherwise, it does not. The oscillation timescale clearly depends on $M_2$, becoming shorter as $M_2$ increases, in agreement with Equation~(\ref{eq:Tsec}). These trends are clearly demonstrated in Figure~\ref{fig:ana}(a) and (b) as well.

Figure~\ref{fig:secular}(c) compares different values of the semimajor axis of the outer perturber $a_2$, fixing its mass and eccentricity at $0.1\,M_\oplus$ and 0.1, respectively. It is clear that the timescale of the eccentricity oscillation strongly depends on $a_2$. The timescale increases with $a_2$: it is about 0.3 Myr for $a_2 = 0.5\,\mathrm{au}$, around 10 Myr for $a_2 = 1\,\mathrm{au}$, and exceeds 100 Myr for $a_2 = 3\,\mathrm{au}$.
This is consistent with the analytical expectations in Section~\ref{sec:ana} and its demonstration in Figure~\ref{fig:ana}(b).

We now examine the case in which multiple gravitationally interacting embryos are placed in the outer orbits. These embryos interact with each other, leading to gravitational scattering and collisions during the evolution. Figure~\ref{fig:secular}(d) shows the result of a simulation in which ten embryos with masses of $0.1\,M_\oplus$ were initially placed between $a=1$ and 2\,au. As in the case of a single perturber, the amplitude of the eccentricity oscillation of the inner planet increases with the initial eccentricity $e_{20}$. However, unlike the single-perturber case, the eccentricities of the outer embryos also evolve over time. Notably, in the $e_{20} = 0.03$ case, the eccentricities of the outer bodies gradually increase and eventually exceed 0.1. As a result, the final eccentricity amplitudes of the inner planet for $e_{20} = 0.03$ and $e_{20} = 0.1$ become comparable (around 0.05--0.1).
Collisions between embryos can also increase their masses, which may slightly enhance the excitation of eccentricities.

Figure~\ref{fig:secular}(e) shows the results when the total mass (i.e., the number) of the outer embryos is varied while keeping the initial eccentricity fixed at 0.1. The eccentricity excitation of the inner super-Earth increases with the total mass of the outer embryos. This is explained by the fact that systems with more mass experience stronger mutual scattering among the embryos, leading to higher eccentricities. At $t = 200\,\mathrm{Myr}$, the maximum eccentricities of the outer embryos reach about 0.2, 0.3, and 0.8 in the simulations with initial 10, 20, and 30 embryos, respectively. This result supports the conclusion from Section~\ref{sec:results} that the orbital stability of inner super-Earths depends on the total mass (and number) of outer embryos. In particular, a larger remaining embryo mass leads to higher eccentricities through mutual scattering, which in turn destabilizes the inner super-Earth system.

Even in systems with multiple embryos, the eccentricity oscillation timescale of the inner super-Earth remains roughly 10 Myr, corresponding to the secular perturbation timescale from each embryo estimated using Equation~(\ref{eq:Tsec}). However, in Figure~\ref{fig:secular}(d)–(f), collisions increase the masses of the embryos, which slightly shortens the oscillation timescale. In addition, it should be noted that in Figure~\ref{fig:secular}(e), the timescale for the eccentricities of the outer embryos to increase due to mutual scattering (which is $>10\,{\rm Myr}$) also appears to be important.

Figure~\ref{fig:secular}(f) shows the results when the total mass and initial eccentricity of the embryos are fixed, but the locations of the embryos are varied. The radial width of the embryo ring is adjusted so that the solid surface density remains constant. As seen in Figure~\ref{fig:ana}(b) and Figure~\ref{fig:secular}(c), the oscillation timescale strongly depends on the semimajor axis of the outer embryos. When the embryos are located beyond $a=3$\,au, the secular timescale exceeds 100\,Myr.

As a side note, the simulations shown in Figure~\ref{fig:secular} consider only one inner super-Earth. If multiple inner super-Earths are in resonance, their mutual precession could alter the way secular forcing from the outer embryos acts on them. It would be interesting to investigate this effect in a separate study; however, since the results of the \textit{N}-body simulations in Section~\ref{sec:results} can be consistently interpreted based on the analysis in Figure~\ref{fig:secular}, the influence of precession induced by resonances is likely to be limited.

\subsection{Summary of conditions for the evolution of observed MMR chains}

In Sections~\ref{sec:ana} and \ref{sec:n-body}, we have investigated how the eccentricities of inner super-Earths can increase due to the secular perturbations from outer embryos. In this subsection, we summarize the findings from Sections~\ref{sec:results} and \ref{sec:secular} and discuss the conditions under which the inner resonance chain becomes unstable over a timescale of 100~Myr.

First, for the eccentricity of the inner super-Earths to increase significantly, it is important that the eccentricities of the outer embryos are large. This is seen in Equation~(\ref{eq:e1max}), where $e_{1,\max}$ is proportional to the initial eccentricity of the outer embryo, $e_{20}$. Figure~\ref{fig:secular}(a), (d), and (e) also confirm that $e_{1,\max}$ depends on $e_2$. Therefore, if the dynamical instability of the inner chain is caused by an increase in eccentricity of the super-Earth, then a large $e_2$ is one of the key conditions for the inner resonances to become unstable.

\begin{figure}[ht!]
\plotone{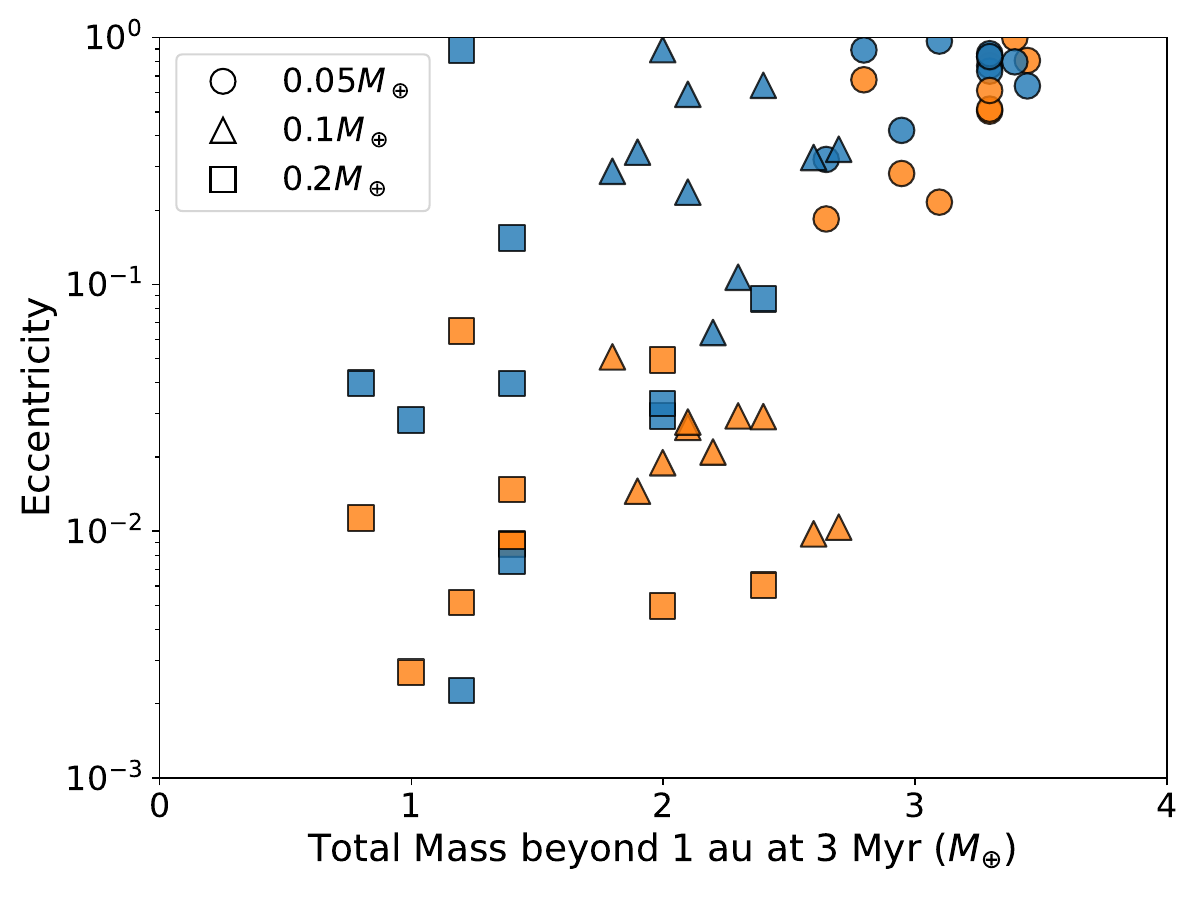}
\caption{Maximum eccentricity of outer embryos with $a > 1\,\mathrm{au}$ in each run of our $N$-body simulations presented in Section~\ref{sec:results}. The horizontal axis shows the total mass of the outer embryos at $t = 3\,\mathrm{Myr}$. Orange symbols correspond to $t = 3\,\mathrm{Myr}$ and blue symbols to $t = 100\,\mathrm{Myr}$. Different symbols indicate different values of $M_{\rm ini}$.}
\label{fig:ecc}
\end{figure}

In our $N$-body simulations presented in Sections~\ref{sec:results} and \ref{sec:secular}, we found that the eccentricities of outer embryos increase due to scattering by super-Earths and mutual gravitational interactions (self-stirring) between embryos. Thus, it is also important to consider under what conditions this excitation becomes efficient. Gravitational stirring among outer embryos is expected to be more effective when their total mass and number are large, because the frequency of mutual scatterings increases in such cases.
Figure~\ref{fig:ecc} shows the results of \textit{N}-body simulations for the slow migration case in Section~\ref{sec:results}. It compares the maximum eccentricities of the outer embryos at $t=3$~Myr and $t=100$~Myr as a function of their total mass at $t=3$~Myr in each simulation. From this figure, we see that in the case of $M_{\rm ini} = 0.05\,M_\oplus$, some outer embryos already exceed an eccentricity of 0.5 at $t = 3\,\mathrm{Myr}$. This increase is considered to be primarily caused by scattering from the super-Earths.  This figure also demonstrates that for $M_{\rm ini} = 0.1\,M_\oplus$, the eccentricities remain below 0.1 at $t = 3\,\mathrm{Myr}$, but are excited to values greater than 0.5 by $t = 100\,\mathrm{Myr}$. We see that the increase in eccentricity over time depends on the total mass. When the total mass is around or above 2~$M_\oplus$, the maximum eccentricity can increase significantly by $t=100$~Myr. On the other hand, when the total mass is below 2~$M_\oplus$, the final eccentricities are generally less than 0.1. In fact, in our \textit{N}-body simulations in Section~\ref{sec:results}, systems with total masses above 2~$M_\oplus$ often experience disruption of the inner resonant chain (e.g., Figure~\ref{fig:3Myr_slow}(d)).

Next, we consider the coefficient $C_{12}$ in Equation~(\ref{eq:e1max}). The secular coefficients in the disturbing function $A_{ij}$ depend on the planetary masses and orbital distances. Fixing a particular inner super-Earth with $M_1$ and $a_1$, the eccentricity excitation becomes weaker when the outer embryo is more distant or has a smaller mass. This is generally consistent with the trends seen in Figure~\ref{fig:ana}(a), though the dependence is somewhat complex. In the parameter range shown in Figure~\ref{fig:ana}(a), where $a_1 = 0.3$~au, $M_1 = 3~M_\oplus$, and $a_2 \lesssim 30$~au with $M_2 \gtrsim 0.05~M_\oplus$, the coefficient is typically greater than 0.01. Therefore, in most cases considered here, this coefficient does not impose a strict condition for inner super-Earths to acquire high eccentricities. 

We now turn to the timescale condition. \citet{2024AJ....168..239D} suggested that the observed evolution of resonance chains occurs over a timescale of roughly 100~Myr. To understand this, we consider the timescale for eccentricity excitation due to secular perturbations, which was shown in Section~\ref{sec:n-body} to be well approximated by Equation~(\ref{eq:Tsec}). This timescale depends on mass $M_2$ and the orbital configuration ($\alpha$ or $a_2$) of the outer embryo. According to Figure~\ref{fig:ana}(b), configurations that result in timescales of $\sim$100~Myr can be identified. For instance, if the perturber is located at $a_2 = 1$~au, its mass should be around $M_2 \sim 0.001$--$0.1~M_\oplus$. If it is at $a_2 = 10$~au, the required mass increases to $M_2 \sim 10$--$100~M_\oplus$. For even larger distances, more massive perturbers are needed.
In our \textit{N}-body simulations, we found that perturbers with $M_2 = 0.01$--$0.1~M_\oplus$ located at $a_2 = 1$--$3$~au, therefore produce eccentricity excitation on the right timescale to reasonably match observations.

Finally, we note that the timescale over which the outer embryos self-excite their eccentricities also plays an important role in the system's overall evolution. As shown in Figure~\ref{fig:ecc}, systems with a total embryo mass of about 2~$M_\oplus$ near $a=$1~au typically reach significant eccentricities over timescales of 10--100~Myr. This is shorter than the 100~Myr condition suggested by observations. In contrast, if the embryos are located farther out, the timescale for eccentricity growth would be longer, potentially inconsistent with the observed timescale.

\section{Conclusions}\label{sec:conclusions}
We have investigated the formation of short-period super-Earths from a ring of embryos initially located near 1 au, and their long-term orbital evolution. We performed $N$-body simulations that include interactions between the planets and the gas disk. The main goal of this study was to clarify which processes can explain the observed distribution of orbital period ratios of super-Earths and their time evolution. We also aimed to identify the cause of orbital instabilities of resonant chains and to place constraints on the conditions that can account for the observed evolution of orbital period ratios. Our main findings are as follows.  

\begin{enumerate}
\item In the simulations, the following process was observed as a pathway for forming super-Earth systems with non-resonant period ratios. Embryos that formed in a ring at $a\simeq$ 1 au grew through mutual collisions in situ. Once they became more massive than about one Earth mass, they underwent inward migration. During migration, neighboring super-Earths were captured into mean-motion resonances, and multiple planets eventually formed a resonant chain. This process is fairly universal and was seen in all simulations.  

\item If the resonant chain becomes dynamically unstable after the disk dispersal, the planets escape from resonance. When this occurs, the final orbital period ratios become larger than 3:2 or 4:3, which better matches the observed distribution of super-Earths. A pileup just outside exact resonances, as seen in observations, was also reproduced. Furthermore, the timescale of instability is longer than the disk dispersal time and can be as long as about 100 Myr. This is consistent with the observed timescale over which resonant period ratios evolve \citep{2024AJ....168..239D}.

\item Whether a resonant chain remains stable or becomes unstable after disk dispersal cannot be explained simply by differences in orbital period ratios, deviations from exact resonance, or the number of planets in the system immediately after formation. Instead, our simulations show that the difference is explained by the presence or absence of embryos that remain on outer orbits beyond $a\simeq$ 1 au. This suggests that the dynamical instability of inner super-Earths is induced by secular perturbations from embryos in outer orbits. While secular perturbations from outer giant planets have often been emphasized in previous studies, our results demonstrate that even small, high-eccentricity embryos can produce similar effects. The dynamical instability of inner resonant chains induced by outer eccentric small planets has also been investigated independently around the same time by \citet{2025A&A_submitted_Goldberg}. The mechanism found in both studies naturally arises in $N$-body simulations and may provide a plausible explanation for why super-Earths in old systems are generally found out of mean-motion resonances.

\item We examined the eccentricity growth of inner super-Earths caused by secular perturbations from outer embryos, using analytic formulas and simple orbital calculations. To break the resonances of inner super-Earths, their eccentricities must be excited to some extent. This requires that the outer embryos have large eccentricities, sufficiently high masses, and not too large semimajor axes (specifically, masses greater than about 0.05 $M_\oplus$ and semimajor axes within 30 au). In addition, since the observed timescale for resonance breaking is about 100 Myr, the secular perturbation timescale (which depends on the masses and semimajor axes of the outer embryos) must also be of this order.  

\item Regarding the large eccentricities of the outer embryos, our simulations showed that embryos remaining beyond 1 au acquire high eccentricities (greater than 0.5) through scattering by forming super-Earths and by their own self-stirring. In the $N$-body simulations, systems with larger total mass in the outer embryos experience stronger self-stirring and are more likely to become unstable. This also explains why systems are more unstable when planets experience slow type I migration compared with rapid migration. With slower migration, more embryos remain on outer orbits, making the system more susceptible to instability.  
\end{enumerate}

\begin{acknowledgments}
We thank Alessandro Morbidelli for his contributions to the previous study that laid the foundation for this work, as well as for his valuable comments on this study. We are also grateful to Max Goldberg and Antoine Petit for stimulating discussions, and to the Observatoire de la Côte d’Azur for their hospitality during our stay in Nice in 2024. We also thank Yanqin Wu for helpful discussions. Finally, we thank the anonymous reviewer for constructive comments that improved the manuscript.
M.O. is supported by the National Natural Science Foundation of China (No. 12273023).
M.K. is supported by JSPS KAKENHI Grant No. JP23K25923.
Numerical computations were in part carried out on PC cluster at the Center for Computational Astrophysics, National Astronomical Observatory of Japan.
\end{acknowledgments}

\appendix

\section{System properties at 3 Myr in the slow migration case}\label{sec:appA}

In Section~\ref{sec:factors}, we examine the factors that influence the dynamical instability of resonant chains. For this purpose, we compare the system properties at $t=3$~Myr for simulations with different initial embryo masses. In the slow migration case (Figure~\ref{fig:3Myr_slow}), the results show that just before disk dispersal, there are no clear differences in the orbital period ratios of super-Earth pairs, in their deviations from exact commensurabilities, or in the number of super-Earths. The main difference appears in the outer region beyond $a \simeq 1$~au, where the remaining embryos show different properties depending on the initial mass. Here, we confirm that similar results are also obtained in the fast migration case.

\begin{figure}[ht!]
\epsscale{0.8}
\plotone{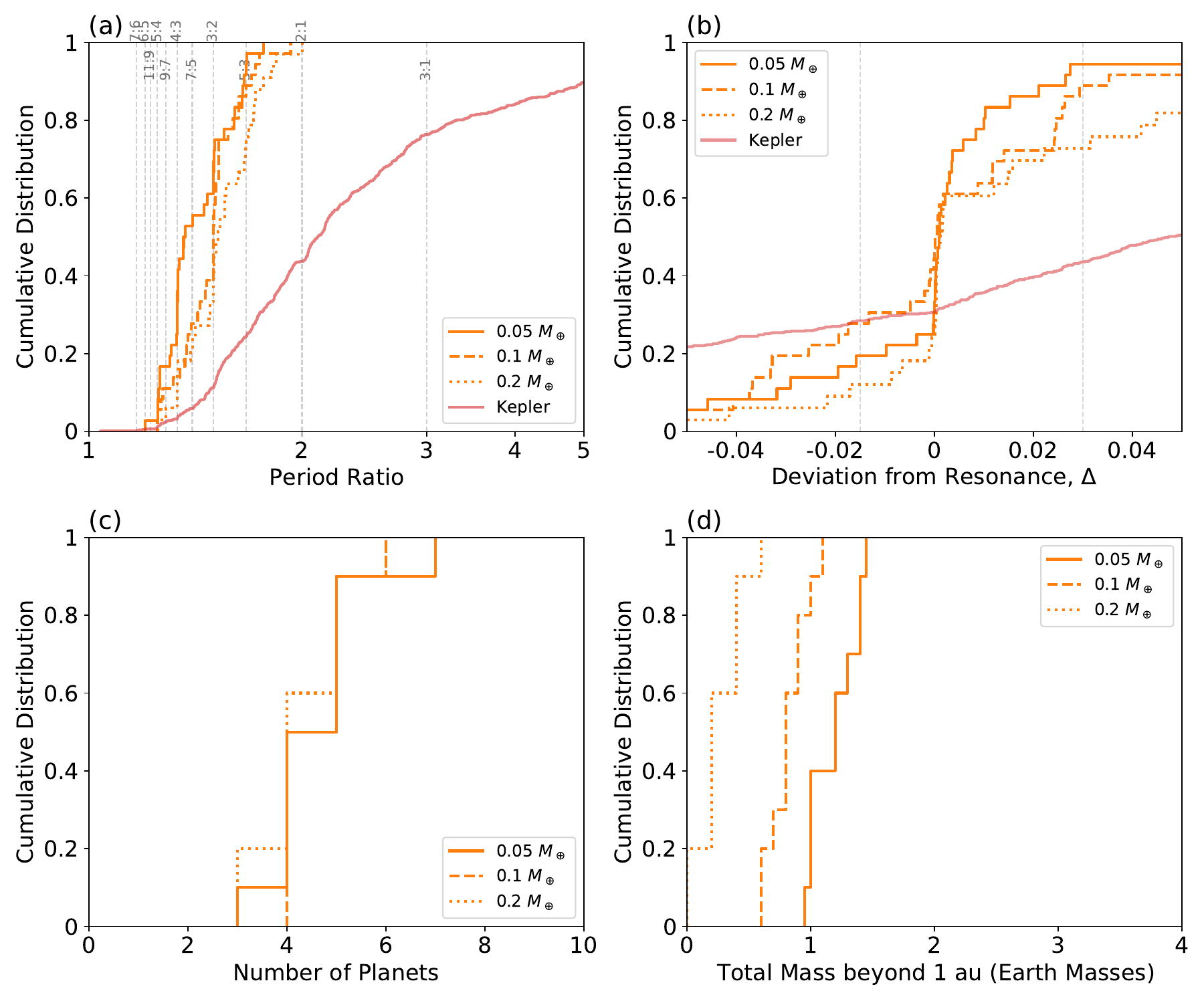}
\caption{Same as Figure~\ref{fig:3Myr_slow}, but showing the results for the fast migration case.}
\label{fig:3Myr_fast}
\end{figure}

Figure~\ref{fig:3Myr_fast} shows the state of the system at $t=3$~Myr in the slow migration case. The overall trends are consistent with those in Figure~\ref{fig:3Myr_slow}. In particular, there are no significant differences in the period ratios or in the number of super-Earths, while the total mass of the embryos remaining on outer orbits differs.

\bibliography{mybib}{}
\bibliographystyle{aasjournalv7}



\end{document}